\documentstyle[12pt]{article}
\begin{document}
\begin{center}
{\Large \bf Centrifugal Force and Ellipticity behaviour}\\[25pt]
{\Large \bf of a slowly rotating}\\[25 pt]
{\Large \bf ultra compact object }\\[0.5truein] 
{\bf Anshu Gupta, Sai Iyer and A.R. Prasanna\\[2pt]
Physical Research Laboratory\\[2pt]
Ahmedabad 380 009, India}\\[30pt]
{\bf \underline{Abstract}}\\[10pt]
\end{center}
{ Using the optical reference geometry approach, we have derived in the
following, a general expression for the ellipticity of a slowly rotating
fluid configuration using Newtonian force balance equation in the
conformally projected absolute 3-space, in the realm of general
relativity. Further with the help of Hartle-Thorne (H-T) metric for a
slowly rotating compact object, we have evaluated the centrifugal force
acting on a fluid element and also evaluated the ellipticity and found
that the centrifugal reversal occurs at around $R/R_s \approx 1.45$, and
the ellipticity maximum at around $R/R_s \approx 2.75$. The result has
been compared with that of Chandrasekhar and Miller which was obtained
in the full 4-spacetime formalism.}

\section{Introduction}
{\noindent One of the enigmatic problems in the context of pulsars is
still the understanding of the internal structure of rotating compact
objects. As normally one considers the object to be slowly rotating,
most of the model calculations have considered the well known
approximate solution of Hartle and Thorne \cite{hartle}, \cite{ht}
as the basic solution for the fluid configuration in the context of 
general relativity. It is generally believed that for any such rotating 
fluid configuration if one considers the `force balance', in the purely 
Newtonian physics one encounters no strange behaviour irrespective of 
the size of the compact object as the two traditional rivals the gravitation 
vis a vis the centrifugal force acting on a fluid element would always be 
opposing each other. However, in the context of general relativity as has 
been shown by Abramowicz and Prasanna \cite{abp} for a sufficiently small 
size object, the centrifugal force acting on a test particle of mass $m_0$ 
in circular orbit outside a static mass $M$ is given by
\begin{equation}
\tilde F_{cfg} = \frac{m_0 \tilde v^2}{r} (1 - \frac{3 M}{r}),
\end{equation}
\\[2pt]
\noindent where $\tilde v$ is the speed of the particle as seen in the
conformally projected 3-space of the optical reference geometry (ACL)
\cite{acl}.  As seen from the above expression the centrifugal force would 
not oppose gravity if the particle is situated at a distance $r \le 3 M$. 
As there could exist ultra compact bodies \cite{vishu} of size $2 M \le R 
\le 3 M$, it would become relevant to consider the effect of such a 
centrifugal force reversal on a fluid element of a possible ultra compact 
rotating configuration.
\\[12pt]
\noindent Another important manifestation of the same result viz, 
introducing Newtonian forces in general relativity is the explanation
of ellipticity maximum for a rotating configuration given by Abramowicz
and Miller \cite{abm}, an effect discovered by Chandrasekhar and Miller
\cite{chnm}. Though the explanation given by Abramowicz and Miller is 
qualitatively viable, quantitatively there appears a difference in the 
location of the ellipticity maximum, which perhaps is due to the fact that 
they considered only the Schwarzschild background geometry, which does not
take into account the influence of rotation of the central body inherently.
\\[12pt]
\noindent In the following, we reexamine the scenario by studying the
possible centrifugal reversal and the ensuing  ellipticity maximum for
a slowly rotating fluid configuration by adopting the Hartle - Thorne
solution which indeed is better suited to study slowly rotating fluid
configuration.
\\[12pt]
\noindent We start with a general axisymmetric, stationary fluid
configuration and introduce a formalism to treat the four forces on a
fluid element  in the 3+1 conformal splitting and then adopting Hartle's
solution as a specific example consider the centrifugal force. Using the
Newtonian principle of force balance equation for a rotating spheroid we
then derive a general expression for the ellipticity and again study its
behaviour for the Hartle solution. We find that the result matches
closer to that of Chandrasekhar and Miller result thus validating the
more general expression derived.

\section{Formalism}
\noindent The equation of motion for a perfect fluid distribution on a
general curved space time
\begin{equation}
ds^2 = g_{i j} dx^i dx^j,
\end{equation}
\noindent are given by
\begin{equation}
(\rho + p) (U^i_{;j} U^j) = - h^{i j} p_{,j}
\end{equation}
\noindent where $\rho$ is the matter density, $p$ the pressure, $U^i$
the four velocity and $h^{i j}$ the projection tensor $(g^{i j} + U^i
U^j)$. This may indeed be expressed as the four force acting on a fluid
element
\begin{eqnarray}
f_i &:=& (\rho + p) (U_{i;j} U^j) + h^j_i p_{,j}\\[8pt]
 &=& (\rho + p) [U^j \partial_j U_i -\frac{1}{2} U^m U^j \partial_i g_{j m}] +
 h^j_i p_{,j}\:\:\:,
\end{eqnarray}
\noindent which when $p = 0$ and $\rho = m_0$, reduces to the well known
four force expression acting on a particle \cite{acl}
\begin{equation}
m_0 f_i \equiv P^j \partial_j P_i -\frac{1}{2} P^j P^m \partial_i g_{j m}.
\end{equation}
\noindent Using the ACL formalism of 3+1 conformal slicing of the space
time
\begin{equation}
ds^2 = dl^2 - g_{0 0} (dt + 2 \omega_{\alpha} dx^{\alpha})^2,
\end{equation}
\noindent with $dl^2$ representing the positive definite metric of the
absolute 3-space ${\tilde g}_{\mu \nu} dx^{\mu} dx^{\nu}$, equation 
(5) may be rewritten as  
\begin{equation}
f_0 = \Phi^{-1} (\rho + p) {\hat U}^\mu \partial_\mu U_0 + h^\mu_0
p_{, \mu}\\[8pt]
\end{equation}
\begin{equation}
\begin{array}{lll}
f_\alpha &=& \Phi^{-1} (\rho + p) \left [ {\hat U}^\mu {\tilde \nabla}_
\mu {\hat U}_\alpha + 2 U^0 {\hat U}^\mu \omega_{\mu \alpha} +
\frac{M_0^2}{2 \Phi} \partial_\alpha \Phi \right ]\\[8pt]
& & + 2 \omega_\alpha f_0 + (h^\mu_\alpha - 2 \omega_\alpha h^\mu_0) p_{, \mu}
\end{array}
\end{equation}
\noindent with
\begin{equation}
\begin{array}{lll}
M_0^2 &=& U_0^2 - {\tilde g}_{\mu \nu} {\hat U}^\mu {\hat U}^\nu\\[8pt]
{\hat U}^\mu &=& \Phi U^\mu\\[8pt]
{\hat U}_{\alpha} &=& {\tilde g}_{\alpha \beta} {\hat U}^{\beta}\\[8pt]
\omega_{\mu \nu} &=& \partial_\mu \omega_\alpha - \partial_\alpha 
\omega_\mu \\[8pt]
g_{0 0} &=& - \Phi\\[8pt]
g_{\mu 0} &=& - 2 \Phi \omega_\mu\\[8pt]
g_{\mu \nu} &=& \Phi {\tilde g}_{\mu \nu} - 4 \Phi \omega_\mu \omega_\nu
\end{array}
\end{equation}
\noindent and ${\tilde \nabla}_\mu$ denotes the covariant derivative in
the absolute 3-space.
\\[12pt]
\noindent As we are interested in slowly rotating fluid configurations,
we shall consider the most general stationary axisymmetric space time
metric as given by \cite{chandra} 
\begin{equation}
ds^2 = -e^{2 \nu} dt^2 + e^{2 \psi} d{\hat \phi}^2 + e^{2 \mu_1} d r^2 +
e^{2 \mu_2} d \theta^2
\end{equation}
\noindent where $\nu, \psi, \mu_1, \mu_2$ and $\omega$ are functions of
$r$ and $\theta$ and $d{\hat \phi} = d\phi - \omega dt$. For this metric
when we have only $V^{\hat \phi} \equiv \frac{d{\hat \phi}}{d t} \ne 0$, 
the normalisation condition yields
\begin{equation}
U^t = \left [ e^{2 \nu} - e^{2 \psi} (\Omega - \omega)^2 \right ]^{-1/2}
\end{equation}
\noindent where $\Omega = \frac{d \phi}{d t} $ is the angular velocity
of the fluid as measured by the stationary observer, and $\omega$ is the
angular velocity acquired by an observer falling freely from infinity.
As this frame $( t, {\hat \phi}, r, \theta)$ is static, the equations (8)
and (9) simplify considerably and then one can calculate the 3-component of
the force as given by
\begin{equation}
f_\alpha  = \Phi^{-1} (\rho + p) \left [ {\hat U}^\mu {\tilde \nabla}_
\mu {\hat U}_\alpha + \frac{M_0^2}{2 \Phi} \partial_\alpha \Phi \right ]
+ h^\mu_\alpha p_{, \mu}
\end{equation}
\noindent which in fact, when zero, gives the equation of hydrodynamic
equilibrium for a rotating fluid configuration. Now given a metric
(approximate or post-Newtonian solution) one can calculate the 
`centrifugal acceleration' term ${\hat U}^\mu {\tilde \nabla}_
\mu {\hat U}_\alpha$ and `gravitational acceleration' term $\frac{M_0^2}{2
\Phi} \partial_\alpha \Phi$, which for the metric (11) yields
\begin{equation}
\begin{array}{lll}
F_{cf} &=& e^{2 \psi + 2 \nu} (\Omega - \omega)^2 (\psi' - \nu') \left [
e^{2 \nu} - e^{2 \psi} (\Omega - \omega)^2 \right ]^{-1}\\[8pt]
F_g &=& e^{2 \nu} \nu',
\end{array}
\end{equation}
\noindent where prime denotes differentiation with respect to $r$.

\subsection*{Ellipticity}

\noindent It is well known that a rotating fluid configuration deviates
from spherical symmetry and depending upon the degree of rotation the
equatorial diameter expands whereas the polar diameter contracts thereby
producing a change in shape. The various equilibrium configurations of
rotating fluids have been well discussed in the literature, and the
sequence goes through Maclaurin spheroids to Jacobi ellipsoids
\cite{ellpfg}. For slowly rotating configuration one can consider the 
Maclaurin spheroid with the ellipticity defined through the usual 
definition of\\[7pt]
\begin{equation}
\epsilon = \frac{1 - (1 - e^2)^{1/2}}{(1 - e^2)^{1/6}},
\end{equation}
\\[7pt]
\noindent $e$ being the eccentricity defined as $e = (1 - b^2/a^2)^{1/2}$,
where $b$ and $a$ are polar and equatorial radii respectively.
\\[12pt]
\noindent Maclaurin had shown that the acceleration due to gravity 
at the equator and pole has the values
\begin{equation}
\begin{array}{lll}
g_{equator} &=& 2 \pi G \rho a \frac{(1 - e^2)^\frac{1}{2}}{e^3}
[sin^{-1}e - e (1 - e^2)^\frac{1}{2}]\\[8pt]
g_{pole} &=& 4 \pi G \rho a \frac{(1 - e^2)^\frac{1}{2}}{e^3} [e -
(1 - e^2)^\frac{1}{2} sin^{-1}e],
\end{array}
\end{equation}
\\[7pt]
\noindent wherein he had considered the possible effects that could
arise due to the internal stresses in the body. However, as we are
looking for a solution in general relativity, wherein the gravitational
field inside the body is described through a metric which is a solution
of Einstein's equations for a perfect fluid distribution, the
gravitational potentials $g_{i j}$ would be incorporating the effects of
all characteristics of the fluid distribution . With this proviso, in
the new language of optical reference geometry it is sufficient to
consider the modified expression for the gravitational and centrifugal
accelerations as given by (14) and use the Newtonian force balance equation
to relate the eccentricity with the acceleration. Thus generalising the
Newtonian equation
\begin{equation}
g_{equator} - a \Omega^2 = g_{pole} (1 - e^2)^\frac{1}{2}.
\end{equation}
\noindent to
\begin{equation}
F_{ge} - F_{cf} = F_{gp} (1 - e^2)^{1/2}
\end{equation}
\noindent and using the force expression as given by
\begin{eqnarray}
(\theta = \pi/2): F_{ge} &=& e^{2\nu_0(r,\pi/2)} \nu_0'{(r,\pi/2)},\\[8pt]
(\theta = 0 ) :F_{gp} &=& e^{2\nu_0(r,0)} \nu_0'{(r,0)},
\end{eqnarray}
\noindent and $F_{cf}$ as in (14), the eccentricity of the configuration
would be given by
\begin{equation}
e^2 = \left(1 - \left[\frac{F_{cf} - F_{ge}}{F_{gp}}\right]^2\right) 
\end{equation}
\noindent and the ellipticity in the limit of slow rotation $e << 1$,
\begin{equation}
\epsilon = \frac{1}{2} e^2.
\end{equation}

\subsection*{Hartle's Solution}

\noindent For a slowly rotating perfect fluid configuration Hartle has
obtained an approximate solution of Einstein's equations given as
\begin{equation}
\begin{array}{lll}
ds^2&=& \left[ - e^{\nu_0} \left( 1 + 2 \left( h_0 + h_2 
P_2 \right) \right) \right] dt^2 + \left( 1 - \frac{\displaystyle 2M}{\displaystyle r} \right)^{-1} \\[8pt]
& &\left[ 1 + 2 
\frac{\displaystyle \left( m_0+m_2P_2 \right)}{\displaystyle (r-2M)} \right] dr^2
+r^2 \left[ 1 + 2  \left( v_2 - h_2 \right) P_2 \right] \\[8pt]
& &\left[ d\theta^2 + \sin^2
\theta (d\phi - \omega dt)^2 \right] + {\cal O}\left(\Omega^3 \right),
\end{array}
\end{equation}
\noindent which represents the rotation as perturbation over the
non-rotating  metric
\begin{equation}
ds^2 = - e^{\nu_0}(r) dt^2 + [1 - 2 M(r)]^{-1} dr^2 + r^2(d\theta^2 +
sin^2 \theta d \phi^2).
\end{equation}
\noindent Expressing the forces as obtained above in terms of the H-T 
potentials one gets
\begin{eqnarray}
F_{cf} &=& r^2 {\bar\omega}^2 (1/r - \nu_0'/2),\\[8pt]
F_{ge} &=& \frac{1}{2} e^{\nu_0} [\nu_0'(1+2h_0-h_2)+2h_0'-h_2']
,\\[8pt]
F_{gp} &=& \frac{1}{2} e^{\nu_0} [\nu_0'(1+2h_0+2h_2)+2h_0'+2 h_2']
,
\end{eqnarray}
\noindent yielding for the ellipticity
\begin{equation}
\epsilon = 3(h_2 + h_2'/\nu_0') + \frac{r^2 {\bar\omega}^2}{e^{\nu_0}}
\left(2/{r \nu_0'} - 1\right).
\end{equation}
}
%\newpage
\section{Results and discussion}
\noindent In the present work we have derived the general expressions
for the centrifugal acceleration $(F_{cf})$ (using ACL formalism) and
ellipticity $(\epsilon)$ (replacing Newtonian acceleration by
relativistic counterparts in the force balance equation (17)). 
Using H-T solution we obtain the values of the centrifugal force and
the ellipticity for a slowly rotating configuration. For the sake of 
comparison we have also calculated the ellipticity as given by 
Chandrasekhar and Miller \cite{chnm}
\begin{equation}
\epsilon_{H-T}(r) = - \frac{3}{2} \left[ \frac{\xi_2(r)}{r} + v_2(r) - h_2(r)
\right].
\end{equation}
\noindent where
\begin{equation}
\xi_2 = 2 p_2^*/(d\nu_0/dr)\:,\:\:\:\:\:\:\:
p_2^* = -h_2 - \frac{1}{3} r^2 e^{-\nu_0} {\bar\omega}^2.\\[8pt]
\end{equation}
\noindent Writing the expressions in dimensionless units:
\begin{equation}
\bar F_{cf} = \frac{F_{cf}}{(G^2 J^2/c^4 R_s^5)},\:\:\:\: \bar \epsilon = \frac{\epsilon}{(G^2 J^2/c^6 R_s^4)},
\end{equation}
\noindent where $J$ is the angular momentum and $R_s (= 2 M)$ is the
Schwarzschild radius, we have evaluated the quantities for a series of
homogeneous configurations with decreasing radii keeping $M$ and $J$
conserved, ${\bar\epsilon}$ denotes our calculations (equation (28))
whereas ${\bar\epsilon}_{H-T}$ denotes the values for corresponding
configurations as obtained using equation (29) in table 1.
\\[12pt]
\noindent Comparing our present result with that of Abramowicz and
Miller, who had obtained the maximum at $R/R_s = 3$, using pure
Schwarzschild geometry, we see that incorporating the effects of
rotation in the geometry (even approximately) improves the result as the
maximum $R/R_s \approx 2.75$ shifts closer to that obtained by 
Chandrasekhar and Miller $R/R_s \approx 2.3$ (fig 2).
\\[12pt]
\noindent Regarding the centrifugal force the general expression
obtained above does show the reversal at $R/R_s \approx 1.45$ and a
maximum at $R/R_s \approx 2.1$ (fig 1). It is interesting to note that 
even after including the effects of fluid distribution in the space time
geometry, the centrifugal force reversal seems to occur closer to the
value as was known in the Schwarzschild geometry. However, as the
ellipticity maximum indicates a possible change in shape of the
configuration, it is to be noted that our expression shows that for a
collapsing configuration, the change occurs earlier ($R/R_s \approx
2.75$) than what had been obtained by Chandrasekhar and Miller  ($R/R_s
\approx 2.3$). As the shape of the body does depend upon the ellipticity
as its value starts decreasing after reaching a maximum, the body would in
principle tend towards a different shape from that of a spheroid. This
could in principle introduce non axisymmetric deformation in the structure
of the body which might generate gravitational radiation. However a full 
significance of the result obtained would become clear only after a more
detailed analysis which takes into account inhomogeneity in the fluid
configuration as well as from more realistic equations of state, than
that of constant density used in the above analysis.

\newpage
\begin{center}
{\bf Figure Captions}\\
\end{center}
\begin{itemize}
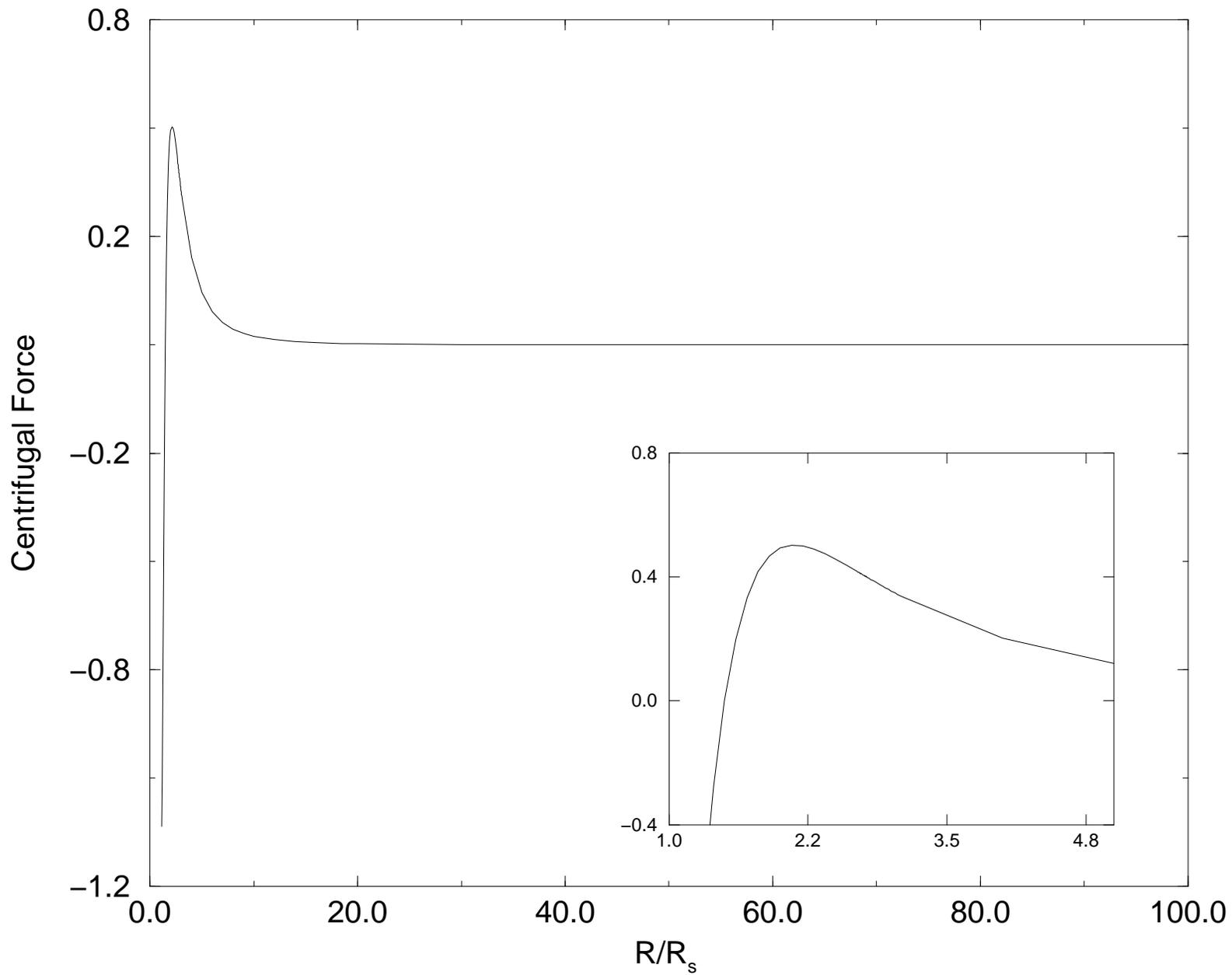
\item[Fig. 1] Plots for centrifugal force ${\bar F}_{cf}$ in units 
$(G^2 J^2/c^4 R_s^5)$ for decreasing values of radius $R$ in terms 
of Schwarzschild radius $R_s$. 
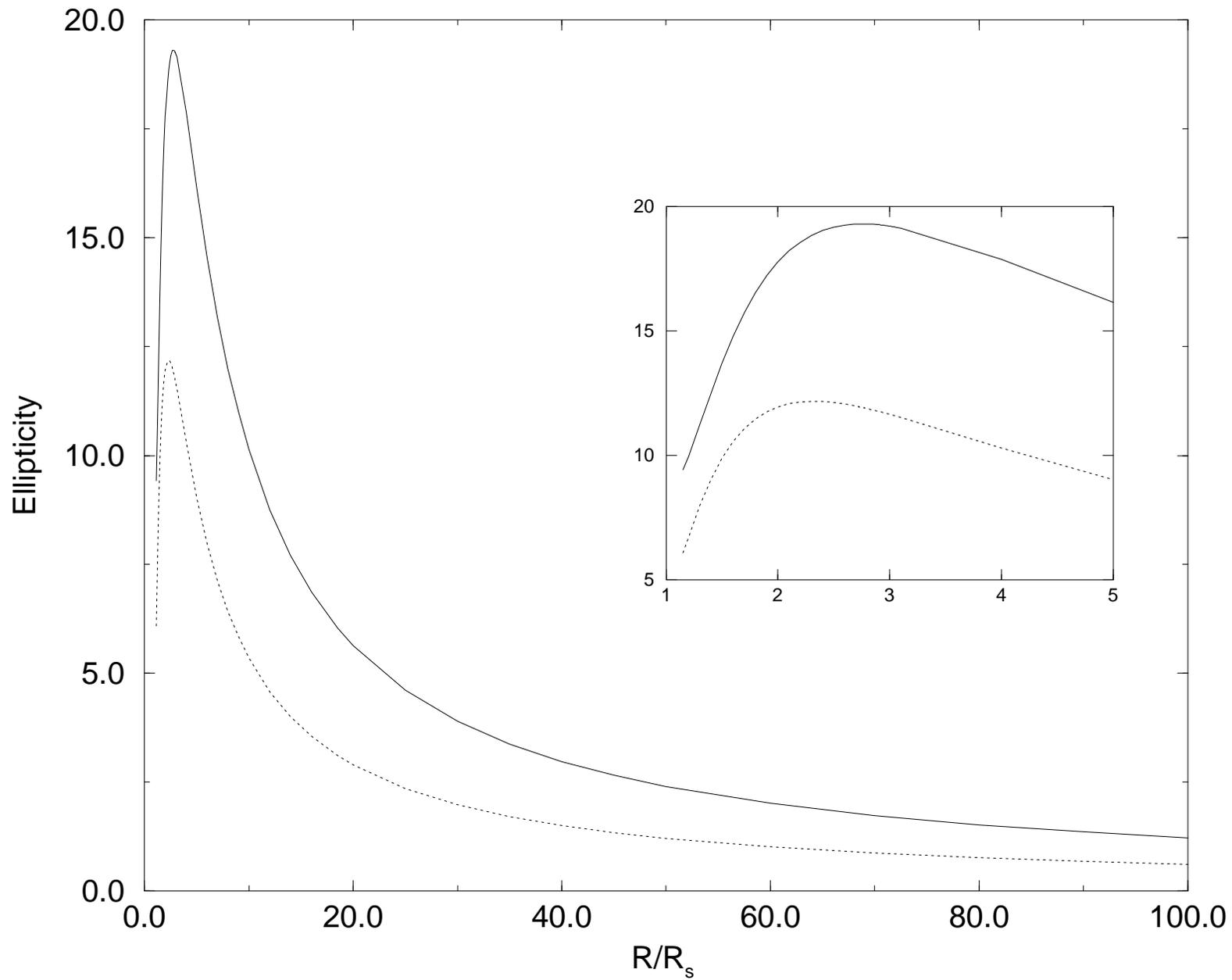
\item[Fig. 2] This shows two curves of ellipticity. The solid line
corresponds to our calculation ${\bar\epsilon}$ and the dotted line
represents the ${\bar\epsilon}_{H-T}$ as used by Chandrasekhar and
Miller. Both the quantities are in units of $(G^2 J^2/c^6 R_s^4)$.
\end{itemize}
\vskip 2cm
\begin{center}
{\bf Table Captions}\\
\end{center}
\begin{itemize}
\item[Table 1] Shows the ellipticity  ${\bar\epsilon}$ (equation
28), ${\bar\epsilon}_{H-T}$ (equation 29) and the centrifugal force 
${\bar F}_{cf}$ (equation 25) (units of these quantities are described
in equation (31)) for a sequence of decreasing radius with conserved 
mass and angular momentum for a homogeneous distribution.
\end{itemize}
\newpage
\begin{table}[htb]
\begin{center}
{\large Table 1}
\\[20pt]
\begin{tabular}{cccc}
\hline
&&&\\
 $R/R_s$&   ${\bar\epsilon}_{H-T}$& ${\bar\epsilon}$& ${\bar F}_{cf}$\\[3mm]
\hline
\hline
&&&\\
  1.125&   5.604732E+0&   9.135673E+0&  -1.105476E+0\\
  1.150&   6.090158E+0&   9.419599E+0&  -1.111393E+0\\
  1.200&   6.728176E+0&   9.973936E+0&  -1.006851E+0\\
  1.300&   7.970848E+0&   1.121079E+1&  -6.367954E-1\\
  1.400&   9.033281E+0&   1.249040E+1&  -2.781407E-1\\
  1.500&   9.893101E+0&   1.370501E+1&   1.036231E-4\\
  1.600&   1.056893E+1&   1.479904E+1&   1.982318E-1\\
  1.700&   1.108810E+1&   1.575131E+1&   3.318464E-1\\
  1.800&   1.147746E+1&   1.656041E+1&   4.172701E-1\\
  1.900&   1.176069E+1&   1.723471E+1&   4.679588E-1\\
  2.000&   1.195771E+1&   1.778676E+1&   4.941394E-1\\
  2.100&   1.208496E+1&   1.823045E+1&   5.033122E-1\\
  2.200&   1.215588E+1&   1.857943E+1&   5.008806E-1\\
  2.300&   1.218139E+1&   1.884639E+1&   4.906997E-1\\
  2.400&   1.217034E+1&   1.904277E+1&   4.755023E-1\\
  2.500&   1.212995E+1&   1.917867E+1&   4.572160E-1\\
  2.600&   1.206606E+1&   1.926295E+1&   4.371926E-1\\
  2.700&   1.198343E+1&   1.930324E+1&   4.163726E-1\\
  2.750&   1.193634E+1&   1.930894E+1&   4.058754E-1\\
  2.800&   1.188595E+1&   1.930607E+1&   3.954040E-1\\
  2.900&   1.177679E+1&   1.927725E+1&   3.747254E-1\\
  3.000&   1.165855E+1&   1.922168E+1&   3.546271E-1\\
  4.000&   1.029998E+1&   1.788030E+1&   2.021015E-1\\
  5.000&   9.029724E+0&   1.615380E+1&   1.209157E-1\\
 10.000&   5.351696E+0&   1.014123E+1&   1.982923E-2\\
 20.000&   2.896627E+0&   5.641617E+0&   2.794844E-3\\
 35.000&   1.710614E+0&   3.370030E+0&   5.475721E-4\\
 50.000&   1.213098E+0&   2.400532E+0&   1.914418E-4\\
 80.000&   7.668072E-1&   1.523602E+0&   4.751822E-5\\
100.000&   6.157617E-1&   1.224927E+0&   2.446291E-5\\[3mm]
\hline
\end{tabular}
\end{center}
\end{table}

\end{document}